\documentclass[12pt,letterpaper]{article}
\usepackage[margin=1in]{geometry}
\usepackage[T1]{fontenc}
\usepackage{times}
\usepackage{setspace}
\usepackage{amsmath}
\usepackage{amssymb}
\usepackage{amsthm}
\usepackage{bm}
\usepackage{booktabs}
\usepackage{graphicx}
\usepackage{tabularx}
\usepackage{array}
\usepackage{url}
\usepackage[authoryear,round]{natbib}
\usepackage[hidelinks]{hyperref}
\usepackage{caption}
\allowdisplaybreaks[2]
\graphicspath{{figs/mor_minimal/}}
\makeatletter
\renewcommand\section{\@startsection{section}{1}{\z@}%
  {-2.5ex \@plus -1ex \@minus -.2ex}{1ex \@plus .2ex}%
  {\normalfont\normalsize\bfseries\MakeUppercase}}
\renewcommand\subsection{\@startsection{subsection}{2}{\z@}%
  {-2ex \@plus -1ex \@minus -.2ex}{0.7ex \@plus .2ex}%
  {\normalfont\normalsize\bfseries}}
\renewcommand\subsubsection{\@startsection{subsubsection}{3}{\z@}%
  {-2ex \@plus -1ex \@minus -.2ex}{0.7ex \@plus .2ex}%
  {\normalfont\normalsize}}
\makeatother
\newcolumntype{Y}{>{\raggedright\arraybackslash}X}
\newtheorem{definition}{Definition}

\newtheorem{proposition}{Proposition}
\newtheorem{corollary}{Corollary}

\usepackage{algorithmicx}
\usepackage{algpseudocode}

\newcounter{readinessalgorithm}
\renewcommand{\thereadinessalgorithm}{\arabic{readinessalgorithm}}

\algrenewcommand{\algorithmicrequire}{\textbf{Input:}}
\algrenewcommand{\algorithmicensure}{\textbf{Output:}}
\algrenewcommand{\alglinenumber}[1]{\scriptsize #1:}
\algrenewcommand{\algorithmicindent}{1em}

\newenvironment{readinessalgorithm}[1][!t]{%
    \par\addvspace{\baselineskip}
    \refstepcounter{readinessalgorithm}
    \setlength{\parindent}{0pt}%
}{%
    \par\addvspace{\baselineskip}
}

\newcommand{\readinessalgorithmcaption}[1]{%
    \noindent
    \textbf{Algorithm~\thereadinessalgorithm: #1}\par
    \vspace{2pt}
    \hrule height 0.4pt
    \vspace{2pt}%
}

\newcommand{\readinessalgorithmend}{%
    \vspace{1.5pt}
    \hrule height 0.4pt%
}

\title{Readiness-Constrained Pre-Islanding Dispatch for Military Microgrids}

\author{Aidan Looney, Nicholas Judson, and Annie Weathers}
\date{}

\begin{document}

\begin{center}
{\Large\bfseries Readiness-Constrained Pre-Islanding Dispatch \\ for Military Microgrids \par}
\medskip
Aidan Looney\textsuperscript{1,2},
Annie Weathers\textsuperscript{2}, and
Nicholas Judson\textsuperscript{2}\par
\smallskip
\begingroup\linespread{1}\selectfont
\textsuperscript{1}John A. Paulson School of Engineering and Applied Sciences,\\
Harvard University, Allston, MA 02134, USA\\
\textsuperscript{2}MIT Lincoln Laboratory, Lexington, MA 02421, USA\\
\par\endgroup
\end{center}

\section*{Abstract}
Military installation energy readiness depends on having sufficient resources to sustain mission-critical loads throughout a prolonged utility outage. This paper develops a method to help operators manage energy resources during normal grid-connected operation while preserving islanding readiness. A robust viability kernel identifies combinations of battery energy, fuel, and load-shifting capability sufficient to sustain critical load under uncertain demand and photovoltaic generation. A probability bound translates stochastic generator start failures and recovery into a deterministic generator count selected to meet a specified availability requirement at islanding. Case studies for a fourteen-day islanding requirement at 0.999 islanding-entry availability show that advance warning of an outage can be used to prepare resources to meet a higher readiness requirement before islanding begins. The case studies considered further demonstrate that economic-only operation and the evaluated fixed battery reserve policy can fail to sustain critical load for the required duration, whereas enforcing the resource-dependent requirement preserves readiness at a modest increase in modeled grid-connected operating cost. These results demonstrate how active resource management can preserve long-duration energy readiness. The framework also helps commanders understand how readiness changes with operating conditions, even when the available resources remain the same.

\begingroup
\setstretch{1}
\noindent\textbf{Keywords:}
Military installation energy readiness, microgrids, viability kernel,
robust optimization, dynamic programming, rolling-horizon dispatch.
\par
\endgroup

\vfill
\begingroup
\footnotesize
\setstretch{1}
\raggedright
\noindent
\textbf{DISTRIBUTION STATEMENT A.}
Approved for public release. Distribution is unlimited.
This material is based upon work supported by the Under
Secretary of War for Research and Engineering under Air Force
Contract No. FA8702-15-D-0001 or FA8702-25-D-B002.
Any opinions, findings, conclusions, or recommendations expressed
in this material are those of the author(s) and do not necessarily
reflect the views of the Under Secretary of War for Research
and Engineering.
\par
\endgroup

\newpage

\newpage

\section{Introduction}
\label{sec:introduction}

Operating a military microgrid requires allocating energy resources to meet electricity demand in real time while preserving the ability to sustain critical missions in the case of an outage in the bulk electric grid. A microgrid is a group of interconnected electrical loads and local energy resources under coordinated control that can operate independently of the utility grid, a mode known as islanding \citep{Broekhoven_Judson2013_military_microgrids}. When a disruption to the bulk electric grid interrupts utility service, the microgrid can disconnect and supply mission-critical facilities and equipment using local generation and storage \citep{NREL2019_microgrid_resilience_defense,judson2012_microgrid}. During normal conditions, operators decide how much power to purchase from the grid, generate locally, and draw from storage. These decisions, collectively called energy dispatch, affect operating costs and the resources available to sustain missions if islanding becomes necessary.

The central problem is to determine which combinations of available resources allow a military microgrid to sustain critical loads for a prescribed period of islanded operation. Readiness depends jointly on battery energy (the battery state of charge represented in megawatt-hours (MWh)), fuel inventory, available generation, demand flexibility, and uncertain demand and photovoltaic solar (PV) generation throughout the outage. These resources must support continued service over the entire mission through operating decisions that adapt as conditions unfold. The operating decision is therefore how to allocate installation energy resources during routine operations while preserving the capability to sustain critical missions for the required islanding duration.

A key question is how to define a readiness policy that reflects military mission requirements. Army Directive 2020-03 assigns senior commanders responsibility for working with mission owners to identify and document the energy requirements of critical missions. When no duration has been specified, the directive calls for planning to sustain energy and water for at least fourteen days \citep{ArmyDirective2020_03}. The Department of War's fiscal year 2023 Annual Energy Performance, Resilience, and Readiness Report also identifies a requirement that all critical-mission energy loads achieve at least 99.9\% availability per fiscal year by fiscal year 2030, with higher availability standards depending on mission criticality \citep{DoW2024EnergyResilience}. Together, these policies motivate a readiness requirement that addresses both endurance and reliability. This study uses a fourteen-day (336-hour) islanding horizon together with a 0.999 availability criterion for the beginning of islanded operation, but the general framework accommodates other mission durations and availability levels.

\subsection{Literature Review}
\label{subsec:lit_rev}

Prior work on military microgrid planning has emphasized resource sizing, system design, and the assessment of resilience during utility outages \citep{marqusee2021_resilience_econ_microgrid,peterson2021_mission_impact_microgrid}. These efforts are supported by quantitative measures of outage performance, including the probability of sustaining critical loads and the extent of service maintained during a disruption \citep{Hutchinson2022_resilience_metrics,Nelson2020_microgrid_resilience}. These analyses principally emphasize system design and outage-performance assessment rather than how routine grid-connected dispatch should preserve sufficient resources for a possible outage. A remaining question is how to connect military endurance requirements to the energy-management decisions made before islanding occurs.

A closely related literature incorporates islanding requirements directly into grid-connected scheduling. Khodaei's formulations evaluate proposed schedules for islanding feasibility and revise them through additional constraints, including a robust formulation that accounts for uncertain load, renewable generation, and outage timing and duration \citep{Khodaei2014_mg_scheduling_time_islanding_constraints,Khodaei2014_microgrid_resilience_scheduling}. Hussain et al.\ similarly revise generator commitments and battery schedules to preserve service to high-priority loads following a grid disruption \citep{hussain2017_feasible_islanding_survivability}. Their subsequent work develops battery state-of-charge constraints to sustain critical loads for a specified number of intervals after an outage, with load and renewable-generation uncertainty incorporated through robust optimization \citep{hussain2018_survivability_constraint_mg_bess}. Zhang et al.\ determine battery reserve thresholds during weather-warning periods and enforce those thresholds in economic operation, accounting for the effect of temperature on battery performance \citep{Zhang2024_proactive_weather_mg}. These studies address islanding readiness through schedule revisions or battery reserve thresholds and are designed to work for relatively short outage durations. The Army's fourteen-day requirement is an endurance problem that requires fundamentally different considerations for resource depletion and uncertainty across days instead of hours. The present work addresses this challenge by pre-computing a readiness boundary over the full modeled resource state for long-duration military islanding.

Viability theory provides a mathematical foundation for characterizing these resource requirements. It studies the states from which a dynamical system can remain within prescribed constraints through admissible operating decisions; the collection of such states is called the viability kernel \citep{aubin2011_viability_main_book}. Numerical methods approximate viable sets by recursively identifying states from which continued constraint satisfaction remains possible \citep{saint1994_approx_viability}. Applied to islanded operation, the robust viability kernel identifies resource states from which an operating policy can sustain critical loads throughout the prescribed mission for every load and PV trajectory in the specified uncertainty set.

\subsection{Proposed Approach and Contributions}

\begin{figure}[htbp]
    \centering
    \includegraphics[width=0.9\textwidth]{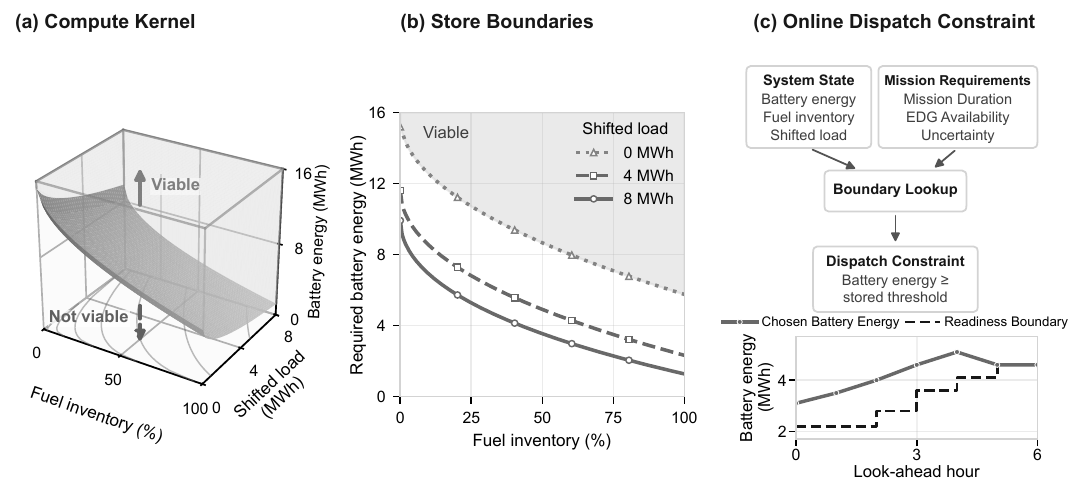}
    \caption{Offline--online architecture for islanding readiness:
    (a) robust viability kernel;
    (b) battery-energy boundary computation;
    (c) boundary query and enforcement in online dispatch.}
    \label{fig:methodology}
\end{figure}

This paper makes two contributions to military microgrid operation. First, it constructs a robust viability kernel that characterizes the combination of available resources sufficient to sustain a long-duration islanded mission under uncertain load and PV generation. A conservative approximation is computed offline and represented through a time-varying battery-energy boundary. Enforcing this boundary in online dispatch permits economically favorable operation during grid-connected operations while preserving sufficient resources to sustain the prescribed islanded mission under the modeled conditions. Figure~\ref{fig:methodology} summarizes the proposed architecture.

Second, the paper characterizes the operational value of advance warning by determining which readiness states can be reached before islanding through charging and load shifting. It also examines how emergency diesel generator (EDG) pre-starting affects generator availability at islanding. This analysis identifies whether a stronger readiness requirement is attainable, the minimum warning time needed, and the associated dispatch cost, thereby quantifying when advance warning makes an otherwise unattainable mission requirement feasible.

\section{Readiness Framework}
\label{sec:kernel}
The framework determines whether the resources available at the start of an outage (i.e., battery energy, fuel, available generators, and load-shifting capability) are sufficient to sustain critical loads for the required outage duration. Because future load and PV generation are uncertain, the method ensures resource requirements still satisfy critical load across a range of possible forecast errors. It then determines the minimum battery energy needed given the other available resources and uncertainty. 

\subsection{Timing and Readiness-State Notation}
Let $\mathcal T:=\{0,1,\ldots,T\}$ denote the discrete state times with uniform time step $\Delta t$. A rolling-horizon optimization initiated at time $t$ plans controls for $s=t,\ldots,t+H-1$ and states through $t+H$. In normal operation, each state time $s\in\{t, \ldots, t+H\}$ is a candidate islanding time. An alert received with lead time $W$ identifies the exact islanding time $I$. The alert then arrives at time $I-W$. When $W\le H$, the known islanding time lies inside the look-ahead dispatch horizon. The post-islanding mission contains $D$ control intervals, indexed by relative offset $h \in \mathcal J_D:=\{0,\ldots,D-1\}$, corresponding to absolute time $s+h$.

The complete readiness state is
\begin{equation}
    z_t:=(e_t,f_t,b_t),
    \qquad
    \chi_t := (z_t,y_t),
    \label{eq:readiness_state}
\end{equation}
where $e_t$ is battery energy in MWh, $f_t$ is fuel inventory in gallons, $b_t$ is shifted load in MWh, and $y_t$ records the declared EDG readiness posture. Operationally, $b_t$ refers to the ability, with advance warning, to serve time-flexible loads (such as building thermal pre-conditioning or electric vehicle charging), shifting those loads to before the outage occurs. $y_t$ refers to the successful generator startups. $z_t$ is the resource state while $\chi_t$ highlights the contribution of generator availability to the kernel as well.

\subsection{Robust Viability Kernel}
\label{subsec:robust_viability_kernel}

Unlike resilience-oriented scheduling models that embed survivability constraints directly in the operating problem \citep{Khodaei2014_mg_scheduling_time_islanding_constraints,Khodaei2014_microgrid_resilience_scheduling,hussain2019_uncertainty_resilient_mg}, the proposed method first characterizes the robust readiness set, then uses the resulting battery-energy boundary to constrain minimum-cost grid-connected dispatch.

For a fixed islanding time $s$, let $\omega_{0:h}:=(\omega_0,\dots,\omega_h)$ denote the load and PV forecast-error history through the hour $h$ after islanding begins. Given uncertainty parameters $\theta$, let $\mathcal U_{s,D}(\theta)$ denote the set of admissible forecast-error trajectories. Each $\omega\in\mathcal U_{s,D}(\theta)$ determines the critical-load trajectory $L_{s+h}(\omega)$ and the PV trajectory $P_{s+h}(\omega)$.

The physical readiness state at the instant of islanding is $z_s$ with the EDG availability handled separately. For the deterministic islanded kernel, the number of EDGs available is treated as a fixed integer $m$. The islanded control policy at offset $h\in\mathcal J_D$ is
\begin{equation}
    \pi_h(\omega_{0:h})
    :=
    \left(
        x^{D,L}_h,\,
        x^{B,L}_h,\,
        x^{PV,L}_h,\,
        x^{PV,B}_h,\,
        x^{PV,C}_h,\,
        \ell_h
    \right).
    \label{eq:islanded_policy}
\end{equation}
All components of $\pi_h$ depend on the observed history ($\omega_{0:h}$). Here, $x^{D,L}_h$ is EDG generation, $x^{B,L}_h$ is battery discharge, $x^{PV,L}_h$ is PV sent directly to load, $x^{PV,B}_h$ is PV sent to the battery, $x^{PV,C}_h$ is curtailed PV, and $\ell_h$ is the allowable critical-load reduction attributable to load shifting before islanding.

The dependence on $\omega_{0:h}$ enforces causality: the control at offset $h$ may depend on the current and previously observed load and PV errors, but not on future realized errors. For conciseness, $\omega$ refers to $\omega_{0:h}$. The following constraints must be satisfied $\forall \ h \in \mathcal{J}_D$ and $\forall \ \omega \in \mathcal{U}_{s,D}(\theta)$:

\begin{subequations}
\begin{align}
    x^{PV,L}_h(\omega)
    +x^{D,L}_h(\omega)
    +x^{B,L}_h(\omega)
    &=
    L_{s+h}(\omega)-\ell_h(\omega),
    \label{eq:islanded_balance}
    \\
    x^{PV,L}_h(\omega)
    +x^{PV,B}_h(\omega)
    +x^{PV,C}_h(\omega)
    &=
    P_{s+h}(\omega),
    \label{eq:islanded_pv_allocation}
    \\
    0\le x^{D,L}_h(\omega)
    &\le m\overline P^G,
    \label{eq:islanded_edg_capacity}
    \\
    0\le x^{B,L}_h(\omega)
    &\le \overline P^{B,-},
    \label{eq:islanded_battery_discharge}
    \\
    0\le x^{PV,B}_h(\omega)
    &\le \overline P^{B,+},
    \label{eq:islanded_battery_charge}
    \\
    0\le \ell_h(\omega)
    &\le \overline P^L,
    \label{eq:islanded_load_shift_power}
    \\
    0\le \ell_h(\omega)
    &\le \beta^L L_{s+h}(\omega),
    \label{eq:islanded_load_shift_bound}
    \\
    x^{PV,L}_h(\omega),\;x^{PV,C}_h(\omega)
    &\ge 0.
    \label{eq:islanded_pv_nonnegativity}
\end{align}
with $\overline P^G$ as the capacity of one EDG, $\overline P^{B,-}$ and $\overline P^{B,+}$, the battery discharge and charge limits, and $\overline P^L$ the shifted-load power limit, and $\beta^L$ is the maximum fraction of critical load that can be shifted in any interval. Shifted-load credit is available only during the first $H^b$ islanded intervals, so $\ell_h(\omega)=0$ for $h\ge H^b$. During islanding, the state satisfies, $\forall \ h \in \{0,\ldots,D\}$:

\begin{equation}
    z_{s+h}(\omega) \in \mathcal Z,
    \
    \mathcal Z := [0,E^{\max}] \times [0,F^{\max}] \times [0,B^{\max}].
    \label{eq:islanded_state_bounds}
\end{equation}
For every $h\in\mathcal J_D$, the state evolves according to
\begin{align}
    e_{s+h+1}(\omega)
    &=
    e_{s+h}(\omega)
    +\eta_c x^{PV,B}_h(\omega)\Delta t
    -\frac{x^{B,L}_h(\omega)}{\eta_d}\Delta t,
    \label{eq:islanded_battery_dynamics}
    \\
    f_{s+h+1}(\omega)
    &=
    f_{s+h}(\omega)
    -\overline r_m\!\left(x^{D,L}_h(\omega)\right)\Delta t,
    \label{eq:islanded_fuel_dynamics}
    \\
    b_{s+h+1}(\omega)
    &=
    \beta_b b_{s+h}(\omega)
    -\frac{\ell_h(\omega)}{\eta_b}\Delta t,
    \label{eq:islanded_loadprep_dynamics}
\end{align}
\end{subequations}
where $\eta_c$, $\eta_d$, and $\eta_b$ are the battery-charging, battery-discharging, and load-shifting efficiency parameters, respectively, and $\beta_b$ defines the holding efficiency for shifted load. The function $\overline r_m$ is a piecewise-linear convex majorant of the manufacturer fuel-burn curve reported in Table~\ref{tab:baseline_parameters}. The available count $m$ bounds total EDG capacity. These dynamics and operating constraints determine which initial resource states can sustain the prescribed islanded mission under all admissible forecast errors. 

\begin{definition}[Robust Viability Kernel]
\label{def:robust_islanded_viability_kernel}
For fixed islanding time $s$, mission duration $D$, uncertainty specification $\theta$, and available EDG count $m$, define
\begin{equation}
    \mathcal K_{s,D,\theta,m} :=\Bigl\{z_s\in\mathcal Z:\; \exists\,\{\pi_h\}_{h\in\mathcal J_D}\textnormal{ such that}\ \eqref{eq:islanded_balance}\textnormal{--}\eqref{eq:islanded_loadprep_dynamics} \textnormal{ hold }\forall\omega\in\mathcal U_{s,D}(\theta) \Bigr\}.
    \label{eq:robust_viability_kernel}
\end{equation}
\end{definition}

Thus, $\mathcal K_{s,D,\theta,m}$ is the set of physical states at the islanding start from which a causal control policy can satisfy the survivability constraints for every admissible load and PV trajectory, conditional on $m$ EDGs remaining available throughout the mission.

\begin{definition}[Readiness Boundary]
\label{def:readiness_boundary}
For fixed $(s,D,\theta,m)$, define the readiness boundary by
\begin{equation}
    \phi_{s,D,\theta,m}(f,b)
    :=
    \inf
    \left\{
        e\in[0,E^{\max}]:
        (e,f,b)\in\mathcal K_{s,D,\theta,m}
    \right\}.
    \label{eq:readiness_boundary}
\end{equation}
If no battery-energy level $e\in[0,E^{\max}]$ makes $(e,f,b)$ viable, then $\phi_{s,D,\theta,m}(f,b):=+\infty$.
\end{definition}

Under the model's compactness and continuity conditions, the kernel is closed, so the infimum in \eqref{eq:readiness_boundary} is attained whenever finite.

The readiness boundary describes how the minimum battery-energy requirement depends on the operating state and mission requirements. The following proposition establishes the direction of these dependencies.

\begin{proposition}[Boundary Monotonicity]
\label{prop:structural_monotonicity}
Greater duration or uncertainty can only reduce viability, whereas additional resources can only improve it. The readiness boundary therefore satisfies
\begin{subequations}
\begingroup
\setlength{\abovedisplayskip}{4pt}
\begin{flalign}
& D_2\ge D_1
\Rightarrow
\phi_{s,D_2,\theta,m}(f,b)
\ge
\phi_{s,D_1,\theta,m}(f,b),
&& \label{eq:phi_duration_monotonicity}\\
& \mathcal U_{s,D}(\theta_1)
\subseteq
\mathcal U_{s,D}(\theta_2)
\Rightarrow
\phi_{s,D,\theta_2,m}(f,b)
\ge
\phi_{s,D,\theta_1,m}(f,b),
&& \label{eq:phi_uncertainty_monotonicity}\\
& f_2\ge f_1
\Rightarrow
\phi_{s,D,\theta,m}(f_2,b)
\le
\phi_{s,D,\theta,m}(f_1,b),
&& \label{eq:phi_f_monotonicity}\\
& b_2\ge b_1
\Rightarrow
\phi_{s,D,\theta,m}(f,b_2)
\le
\phi_{s,D,\theta,m}(f,b_1),
&& \label{eq:phi_b_monotonicity}\\
& m_2\ge m_1
\Rightarrow
\phi_{s,D,\theta,m_2}(f,b)
\le
\phi_{s,D,\theta,m_1}(f,b).
&& \label{eq:phi_m_monotonicity}
\end{flalign}
\endgroup
Also, each viable battery-energy slice is upward closed. For fixed
$(f,b)$ and $0 \leq e_1 \leq e_2 \leq E^{\max}$,
\begin{equation}
    (e_1,f,b)\in\mathcal K_{s,D,\theta,m}
    \Longrightarrow
    (e_2,f,b)\in\mathcal K_{s,D,\theta,m}.
    \label{eq:kernel_energy_upper_set}
\end{equation}
\end{subequations}
\end{proposition}

These properties follow by reusing a feasible policy under a shorter mission, a smaller uncertainty set, or a state with additional resources. Extra battery energy can be left unused because discharge may be withheld and PV may be curtailed. Likewise, additional fuel or shifted-load credit need not be consumed. For $m_2\ge m_1$, additional EDGs may remain idle, and the fuel-burn model satisfies $\overline r_{m_2}(p)\le\overline r_{m_1}(p)$ for $0\le p\le m_1\overline P^G$. Thus, a policy feasible with $m_1$ available EDGs remains feasible with $m_2$.

A state is viable if and only if its battery energy meets or exceeds the readiness boundary:
\begin{equation}
    (e,f,b)\in\mathcal K_{s,D,\theta,m}
    \quad\Longleftrightarrow\quad
    e\ge\phi_{s,D,\theta,m}(f,b).
    \label{eq:boundary_membership}
\end{equation}
The readiness boundary therefore converts the robust multi-hour islanded survivability requirement into a scalar battery-energy threshold, conditioned on the rest of the resource state. For fixed $(s,D,\theta,m)$, the boundary partitions the $(f,b)$-space into three operational regimes. A resource state is \emph{resource-limited} when $\phi=+\infty$, \emph{battery-relevant} when $0<\phi\le E^{\max}$, and \emph{resource-abundant} when $\phi=0$. The boundary is computed offline using the carried-budget Bellman recursion described below.

\subsection{Uncertainty Characterization}
\label{subsec:uncertainty_characterization}

The model represents the hourly forecast error and how long forecast errors can persist throughout islanding. It achieves this with separate, centered, forecast-relative budget uncertainty sets for critical load and PV generation. Let $\widehat L_{s+h}$ and $\widehat P_{s+h}$ denote the forecasts at islanded offset $h\in\mathcal J_D$. The true output is the product of the forecast and the relative forecast error parameter $\xi$:
\begin{equation}
    L_{s+h}
    =
    \widehat L_{s+h}(1+\xi^L_h),
    \qquad
    P_{s+h}
    =
    \widehat P_{s+h}(1+\xi^{PV}_h).
    \label{eq:forecast_plus_error}
\end{equation}

Following the budgeted-uncertainty construction in \citep{Bertsimas2004_price_robustness}, define
\begin{equation}
    \Xi^j(\rho^j,\Gamma^j) :=\bigl\{\xi^j\in\mathbb R^D:\; \|\xi^j\|_\infty\le \rho^j,\quad \|\xi^j\|_1\le \Gamma^j\bigr\},
    \qquad j\in\{L,PV\}.
    \label{eq:load_pv_uncertainty}
\end{equation}
The parameter $\rho^j$ limits each single-interval relative error, while $\Gamma^j$ limits the total absolute relative error over the full islanded period. At $\Gamma^j=D\rho^j$, the budget constraint is redundant. Collect these parameters in $\theta:=(\rho^L,\Gamma^L,\rho^{PV},\Gamma^{PV})$ and define the joint uncertainty set as $\mathcal U_{s,D}(\theta):=\Xi^L(\rho^L,\Gamma^L)\times\Xi^{PV}(\rho^{PV},\Gamma^{PV})$.

For the backward recursion, it suffices to consider true load higher than forecasted and true PV generation lower than forecasted. Lower load and additional PV cannot make islanded operation less feasible because resource use can be reduced and excess PV curtailed. We therefore use nonnegative deviations $\zeta_h^L:=\max\{\xi_h^L,0\}$ and $\zeta_h^{PV}:=\max\{-\xi_h^{PV},0\}$ for the carried-budget Bellman recursion.

\subsection{Carried-Budget Bellman Recursion}
\label{subsec:remaining_budget_bellman}
The viability kernel uses a Bellman recursion by working backward from the end of the required outage period. At each hour, the method determines the minimum resources needed to ensure the mission can be completed under the modeled uncertainty. As the calculation moves backward, it tracks how much load and PV uncertainty could still occur during the remaining outage. 

Fix $(s,D,\theta,m)$ throughout this subsection. To preserve the mission-wide uncertainty limits during backward induction, the offline dynamic program carries the remaining load and PV uncertainty budgets as information states. For $j\in\{L,PV\}$, define the budget remaining before islanded interval $h$ by
\begin{equation}
    g_h^j
    :=
    \Gamma^j-\sum_{n=0}^{h-1}\zeta_n^j,
    \qquad
    g_0^j=\Gamma^j,
    \label{eq:remaining_budget_definition}
\end{equation}
with the update $g_{h+1}^j=g_h^j-\zeta_h^j$. Given $g_h^j$, the current adverse error lies in $\Delta^j(g_h^j):=[0,\min\{\rho^j,g_h^j\}]$.
Thus, a larger error in the current interval leaves a smaller uncertainty budget for the remaining mission.

Let $\mathbf g:=(g^L,g^{PV})$ and $\boldsymbol\zeta:=(\zeta^L,\zeta^{PV})$, and define the joint set of possible forecast errors for the current hour as $\boldsymbol\Delta(\mathbf g):=\Delta^L(g^L)\times\Delta^{PV}(g^{PV})$.
Let $u$ denote the one-hour islanded action, $\mathcal C_{s+h}(z,\boldsymbol\zeta;m)$ the corresponding feasible-action set, and $F_{s+h}(z,u,\boldsymbol\zeta;m)$ the state transition defined by \eqref{eq:islanded_battery_dynamics}--\eqref{eq:islanded_loadprep_dynamics}. At interval $h$, the physical state $z$ and remaining budgets $\mathbf g$ are known before $\boldsymbol\zeta$ is observed. The action is then chosen after observing the current error, and the next information state is $\mathbf g-\boldsymbol\zeta$.

Because no terminal reserve is imposed beyond the physical state bounds, the terminal set is $\mathcal Z$ in \eqref{eq:islanded_state_bounds}. Let $\mathcal K^{\rm DP}_{s,h}(\mathbf g;m)$ denote the physical states from which the remaining $D-h$ islanded intervals can be completed while satisfying the remaining uncertainty budgets $\mathbf g$. For $h=D-1,\ldots,0$, the exact backward recursion is

\begin{equation}
\begin{aligned}
\mathcal K^{\rm DP}_{s,h}(\mathbf g;m)
:=
\bigl\{z:\;&
\forall\boldsymbol\zeta\in\boldsymbol\Delta(\mathbf g),
\quad\exists u\in\mathcal C_{s+h}(z,\boldsymbol\zeta;m)\\[-1mm]
&\text{such that }F_{s+h}(z,u,\boldsymbol\zeta;m)
\in
\mathcal K^{\rm DP}_{s,h+1}
(\mathbf g-\boldsymbol\zeta;m)
\bigr\}.
\end{aligned}
\label{eq:carried_budget_bellman_recursion}
\end{equation}
The corresponding remaining-mission readiness boundary is
\begin{equation}
\begin{aligned}
\phi^{\rm DP}_{s,h}(f,b,\mathbf g;m)
:=
\inf\bigl\{
e\in[0,E^{\max}] :\,
(e,f,b)\in
\mathcal K^{\rm DP}_{s,h}(\mathbf g;m)
\bigr\},
\end{aligned}
\label{eq:stage_readiness_boundary}
\end{equation}
with value $+\infty$ when the set is empty. Larger remaining uncertainty budgets allow more adverse future errors and therefore cannot reduce the minimum battery-energy requirement. Hence, if $\mathbf g_2\ge\mathbf g_1$ componentwise,
\begin{equation}
    \phi^{\rm DP}_{s,h}(f,b,\mathbf g_2;m)
    \ge
    \phi^{\rm DP}_{s,h}(f,b,\mathbf g_1;m).
    \label{eq:budget_monotonicity}
\end{equation}
At the initial budgets, the recursion returns $\phi^{\rm DP}_{s,0}\bigl(f,b,(\Gamma^L,\Gamma^{PV});m\bigr)$.

\begin{proposition}[Bellman representation and numerical conservatism]
\label{prop:bellman_certificate}
For fixed $(s,D,\theta,m)$, the trajectory-based viability kernel and the exact carried-budget Bellman recursion satisfy
\begin{subequations}
\begin{equation}
    \mathcal K_{s,D,\theta,m}
    =
    \mathcal K^{\rm DP}_{s,0}
    \bigl((\Gamma^L,\Gamma^{PV});m\bigr).
    \label{eq:kernel_dp_equivalence}
\end{equation}
Let $\widehat\phi_{s,D,\theta,m}(f,b)$ denote the numerical battery-energy boundary returned by Algorithm~\ref{alg:offline_boundary} at the initial uncertainty budgets $(\Gamma^L,\Gamma^{PV})$. The construction in the Appendix~\ref{sec:offline_algorithm} yields a conservative approximation: for every $(e,f,b)\in\mathcal Z$,
\begin{equation}
    e\ge\widehat\phi_{s,D,\theta,m}(f,b)
    \quad\Longrightarrow\quad
    (e,f,b)\in\mathcal K_{s,D,\theta,m}.
    \label{eq:numerical_boundary_sufficiency}
\end{equation}
\end{subequations}
\end{proposition}
The proof is given in the Appendix.

\subsection{EDG Availability at Islanding}
\label{subsec:probabilistic_edg_availability}

The EDG model determines the generator count credited by the viability kernel at a specified reliability level. Available generators at islanding start are assumed to remain available for the islanded mission. Let $N$ be the total number of EDGs and let $y$ units be pre-started $W$ periods before islanding, with the remaining $N-y$ units starting at islanding. Each EDG starts successfully with probability $p_s$; after a failed start, recovery occurs independently in each period with probability $\mu$. A pre-started EDG has availability probability \mbox{$p^{\rm prep}(W)=1-(1-p_s)(1-\mu)^W$} at islanding. Assuming independent outcomes across units, the available count is the sum of independent binomial variables:
\begin{equation}
    M_I(W,y)=\mathrm{Binomial}\!\left(y,p^{\rm prep}(W)\right)+\mathrm{Binomial}\!\left(N-y,p_s\right).
    \label{eq:M_s_decomp}
\end{equation}
For a candidate islanding time $s$, write this count as $M_s(y_s)$, where the declared posture $y_s$ records the preparation lead and pre-started count.

For a physical state $z_s$, define the minimum number of EDGs needed for viability as
\begin{equation}
    m^{\rm req}_{s,D,\theta}(z_s)
    :=
    \min\left\{
        m\in\{0,\ldots,N\}:
        z_s\in\mathcal K_{s,D,\theta,m}
    \right\},
    \label{eq:required_edg_count}
\end{equation}
with $m^{\rm req}_{s,D,\theta}(z_s)=+\infty$ if no such $m$ exists.

Let $q\in(0,1)$ denote the required probability that enough
EDGs are available at islanding. The $q$-reliable islanded entry target is
\begin{equation}
    \mathcal A^{q,\rm entry}_{s,D,\theta}(y_s)
    :=
    \left\{
        z_s:
        \mathbb P\!\left(
            M_s(y_s)\ge m^{\rm req}_{s,D,\theta}(z_s)
        \right)\ge q
    \right\}.
    \label{eq:q_reliable_kernel}
\end{equation}
The associated EDG count is
\begin{equation}
    m_q(y_s)
    :=
    \max
    \left\{
        m\in\{0,\ldots,N\}:
        \mathbb P(M_s(y_s)\ge m)\ge q
    \right\}.
    \label{eq:certified_edg_count}
\end{equation}
That is, $m_q(y_s)$ is the largest number of EDGs available at islanding with probability at least $q$. The parameter $q$ is an operator-selected per-event islanded-entry criterion, taken from the legal annual-availability metric~\citep{DoW2024EnergyResilience}, excluding failures from already-running generators during the actual islanded operation.

\begin{corollary}[Reliability-based boundary selection]
\label{prop:q_reliable_reduction}
By Proposition \ref{prop:structural_monotonicity}, the viability kernels are nested in available EDG count. Therefore, for any declared EDG state $y_s$,
\begin{subequations}
\begin{equation}
    \mathcal A^{q,\mathrm{entry}}_{s,D,\theta}(y_s)
    =
    \mathcal K_{s,D,\theta,m_q(y_s)}.
    \label{eq:q_reliable_reduction}
\end{equation}
Thus, $z_s=(e_s,f_s,b_s)$ satisfies the $q$-reliable entry requirement if and only if
\begin{equation}
    e_s
    \ge
    \phi_{s,D,\theta,m_q(y_s)}(f_s,b_s).
    \label{eq:q_reliable_boundary}
\end{equation}
\end{subequations}
\end{corollary}

\begin{proof}
By kernel nesting, a state requiring $m^{\rm req}_{s,D,\theta}(z_s)$ EDGs is viable whenever $M_s(y_s)\ge m^{\rm req}_{s,D,\theta}(z_s)$. By the definition of $m_q(y_s)$, this event has probability at least $q$ if and only if $m^{\rm req}_{s,D,\theta}(z_s)\le m_q(y_s)$.
This is equivalent to
$z_s\in\mathcal K_{s,D,\theta,m_q(y_s)}$.
The boundary condition follows directly from the definition of
$\phi_{s,D,\theta,m}$.
\end{proof}

This reduction incorporates stochastic EDG entry availability into the deterministic viability framework through the count $m_q(y_s)$, which determines the battery, fuel, and load shifting resources required for readiness. 

\section{Readiness-Constrained Online Operation}
\label{sec:online_operation}

We define normal and alert readiness standards as
\begin{subequations}
\begin{equation}
    \Theta^N:=(D^N,\theta^N,q^N), \quad \Theta^A:=(D^A,\theta^A,q^A).
\end{equation}
The alert standard keeps the mission duration fixed but strictly increases the uncertainty and entry-reliability requirements:
\begin{equation}
    D^A=D^N, \quad
    \theta^A \succ \theta^N, \quad
    q^A>q^N.
    \label{eq:alert_dominates_normal}
\end{equation}
\end{subequations}
In the case study, the required islanded mission duration is 14 days under both normal and alert conditions, so $D^N=D^A=336$ h. A warning raises the readiness requirement by increasing the load and PV uncertainty allocations, as well as the required entry-reliability level. When the warning identifies the islanding time $I$, the dispatcher uses the available lead time to bring the microgrid into the alert-readiness target set through actions like charging the battery, load shifting, or pre-starting generators. This alert-readiness target set is defined as the following,
\begin{equation}
    \mathcal A_I^A:=
    \left\{
    \chi_I: e_I \ge \phi_{I,D^A, \theta^A, m_{q^A}(y_I)}(f_I,b_I)
    \right\}.
    \label{eq:terminal_alert_target}
\end{equation}
Thus, by the start of islanding, the battery energy must meet or exceed the alert readiness boundary evaluated at the resource state at that time.

\subsection{Online Dispatch}
\label{sec:online_dispatch}

The online dispatch uses the precomputed readiness boundaries to ensure that sufficient resources are available during an outage. During normal operation, the controller minimizes cost while maintaining readiness. If warning is received of an impending outage, it moves the dispatch to a higher readiness requirement. Let $\sigma_t\in\{N,A\}$ denote the active operating mode: normal or alert. Normal mode applies until a warning ($W$) identifies islanding time $I$, at which point alert mode applies from $I-W$ to $I$. Because EDGs do not serve grid-connected load in this model, fuel is unchanged over the online horizon and the measured inventory $f_t$ is passed directly to the boundary.

Within each online solve, subscripts $s$ denote absolute time; $e_t$ and $b_t$ are fixed to the measured battery-energy and shifted-load components of $z_t$, respectively. For $s=t,\ldots,t+H-1$, let $x^{G,L}_s,x^{G,B}_s,x^{G,b}_s$ denote grid power sent to load, the battery, and shifted-load production, respectively. Define the analogous PV allocations $x^{PV,L}_s,x^{PV,B}_s,x^{PV,b}_s$, along with PV export $x^{PV,G}_s$, curtailment $x^{PV,C}_s$, and battery discharge $x^{B,L}_s$. For conciseness, write $p^{B,+}_s:=x^{G,B}_s+x^{PV,B}_s$ and $p^{b,+}_s:=x^{G,b}_s+x^{PV,b}_s$. In either operating mode, the online problem is the following with $\lambda_s$ as the electricity price at time $s$.
\begin{subequations}
\begin{equation}
    J_t^{\sigma_t}
    =
    \min
        \sum_{s=t}^{t+H-1}\bigg(\lambda_{s} \left(x^{G,L}_s+x^{G,B}_s+x^{G,b}_s-x^{PV,G}_s\right) +c^B\left(p^{B,+}_s+x^{B,L}_s\right) +c^b p^{b,+}_s\bigg)\Delta t,
    \label{eq:online_dispatch_objective}
\end{equation}
subject to, $\forall \ s\in\{t,\ldots,t+H-1\}$,
\begin{align}
    x^{G,L}_s+x^{PV,L}_s+x^{B,L}_s
    &=\frac{1}{\beta^f}\widehat L_{s},
    \label{eq:online_load_balance}\\
    x^{PV,L}_s+x^{PV,B}_s+x^{PV,b}_s+x^{PV,G}_s+x^{PV,C}_s
    &=\widehat P_{s},
    \label{eq:online_pv_balance}\\
    e_{s+1}
    &=e_s+\eta_c p^{B,+}_s\Delta t
      -\frac{x^{B,L}_s}{\eta_d}\Delta t,
    \label{eq:online_battery_dynamics}\\
    b_{s+1}
    &=\beta_b b_s+p^{b,+}_s\Delta t,
    \label{eq:online_prepared_dynamics}\\
    p^{B,+}_s
    &\le\overline P^{B,+}, \quad
    x^{B,L}_s\le\overline P^{B,-},
    \label{eq:online_battery_power_limit}\\
    p^{b,+}_s
    &\le \mathbb{I}\{I-W\le s<I\}\,\overline P^b,
    \label{eq:online_prepared_limit}\\
    0\le e_s
    &\le E^{\max},\quad 
    0\le b_s\le B^{\max}.
    \label{eq:online_prepared_bounds}
\end{align}
All allocation variables are nonnegative. The indicator $\mathbb{I}\{I-W\le s<I\}$ equals one during the warning interval and zero otherwise, including normal operation before a warning is received. Thus, load shifting is available only after warning and before islanding. The coefficients $c^B$ and $c^b$ impose small positive battery and load-shifting costs, and $\beta^f$ is the fraction of load that is critical.

At a rolling solve initiated at $t$, these floors are applied to each modeled state time for which islanding remains a candidate,
\begin{equation}
    e_s\ge \hat{\phi}_{s, D^N, \theta^N, m_{s}^N}(f^N,0),
    \qquad s\in\{t,\ldots,t+H\}.
    \label{eq:normal_readiness_floors}
\end{equation}
When no exact alert target is active, the terminal condition
\begin{equation}
    e_{t+H}\ge
    \max_{s\in\{t,\ldots,t+H\}}
    \hat{\phi}_{s, D^N, \theta^N, m_{s}^N}(f^N,0)
    \label{eq:normal_continuation_floor}
\end{equation}
prevents the finite dispatch horizon from satisfying an intermediate floor and then depleting the battery in its final modeled interval.

In alert mode, when $t\le I\le t+H$, the rolling horizon is truncated at $I$: controls and costs use $s=t,\ldots,I-1$, while state bounds and normal readiness floors use $s=t,\ldots,I$. At the measured fuel inventory, the stored boundary is evaluated at each shifted-load grid point and connected by adjacent chords. Because the stored boundary is convex in shifted load at fixed fuel, this piecewise-linear representation, denoted $\widehat \phi^{\mathrm on}$, weakly overestimates the stored boundary between grid points. The alert constraint is
\begin{equation}
    e_I
    \ge
    \widehat\phi^{\rm on}_{I,D^A,\theta^A,m_I^A}
    \left(f_t,b_I\right).
    \label{eq:online_alert_boundary}
\end{equation}
\end{subequations}
Alert operation uses the same economic objective but adds the exact-time readiness constraint \eqref{eq:online_alert_boundary} to the normal readiness floors. Thus, the long-duration islanded problem enters the online dispatch only through the queried boundary.

\subsection{Warning-Triggered Alert Reachability}
\label{sec:value}

Advance warning gives operators time to prepare the microgrid for an outage, but meeting the higher alert standard depends on its current condition and how much warning is available. For an initial state $\chi_{t_0}$, let $\mathcal R_W(\chi_{t_0})$ denote the states attainable at islanding time $I$ when the specified least-cost normal policy is followed before $I-W$, followed by feasible grid-connected preparation under the online operating constraints. The switch to alert-standard readiness is reachable with warning $W$ if and only if $\mathcal R_W(\chi_{t_0})\cap\mathcal A_I^A\ne\varnothing$. 

The best attainable readiness margin is
\begin{equation}
    \mu_W^A
    :=
    \max_{\chi_I\in\mathcal R_W(\chi_{t_0})}
    \left[
        e_I
        -
        \phi_{I,D^A,\theta^A,m_I^A}(f_I,b_I)
    \right].
    \label{eq:alert_readiness_margin}
\end{equation}
Alert readiness is attainable if and only if $\mu_W^A\ge0$. The minimum sufficient warning lead is the smallest $W$ satisfying this condition. For a fixed alert target ($\Theta^A$), longer warning permits the shorter-warning preparation strategy, so $\mu_W^A$ is nondecreasing in $W$.

For warning leads that reach the alert target, we define the incremental preparation cost as $\Delta C^A(W):=J_W^A-J^N$. Here, $J_W^A$ is the realized grid-connected cost under the least-cost normal policy before warning and the least-cost alert-feasible policy thereafter, while $J^N$ is the cost of continued normal operation over the same accounting interval. These two metrics are assessed in the simulations below.

\section{Case Study and Results}
\label{sec:results}

The simulations use hourly data and the parameters in Table~\ref{tab:baseline_parameters}. In both normal and alert operation, $\Gamma^j/\rho^j=300$ so the alert standard increases the error bounds while preserving the normalized uncertainty budget. The illustrative total-load shape combines normalized ComStock profiles for a medium-office, large-hotel, warehouse, and full-service-restaurant. These are aggregated into a single profile with weights 0.40, 0.30, 0.20, and 0.10, respectively, scaling the synthetic load profile to a 13.35-MW annual mean \citep{NRELOEDIEndUseProfiles}. Critical load is 50\% of that load profile, and noncritical load is not served during the islanded period.

PV generation is taken from PVWatts V8 hourly output for 35.1415$^\circ$N, 79.0080$^\circ$W, corresponding to Fort Bragg, NC. PV parameters include a 12-MW$_{\rm dc}$ standard-module, fixed-open-rack system, 14\% losses, south-facing tilt at local latitude, a 1.1 dc/ac ratio, 96\% inverter efficiency, and NSRDB typical-meteorological-year weather \citep{NRELPVWattsV8}. Electricity prices from June 9, 2025 through June 9, 2026 corresponding to the Fort Bragg, NC region are matched by month, day, and hour to the synthetic load/PV data \citep{PJMDataMiner2}. The primary evaluated operating window begins July 29 and extends for a week, containing 168 candidate islanding hours. An autumn sensitivity on October 28, 2025 was selected for Figure~\ref{fig:uncertainty_effects} to assess different seasonal effects on load and PV.

The numerical boundary uses three points per budget interval, an adaptively refined fuel grid with 100-gal increments, and five shifted load points, and this boundary is conservative by Proposition~\ref{prop:bellman_certificate}. Computations were performed on an Apple M4 computer with Gurobi to solve the offline boundary and online dispatch. The 198 boundary libraries required a median runtime of 29.1 min per library and a total of 53.3 hours. The stored boundaries are reused across dispatch runs and querying them required approximately 0.18 ms of median additional runtime. The computationally expensive viability calculations are performed offline, allowing dispatch decisions to account for the full islanding mission with little additional online computation.

\begin{table}[tbp]
\caption{Baseline Simulation Parameters}
\label{tab:baseline_parameters}
\centering
\linespread{1}\selectfont
\small
\setlength{\tabcolsep}{3pt}
\renewcommand{\arraystretch}{1.08}
\begin{tabularx}{\columnwidth}{@{}>{\raggedright\arraybackslash}p{0.2\columnwidth}Y@{}}
\toprule
Group & Baseline values \\
\midrule

Installed resources
& PV: 12 MW; battery: 16 MWh/4 MW with
$\eta_c=\eta_d=0.902$; $F^{\max}=198{,}000$ gal; and
$13\times0.80$-MW EDGs. \\

Mission requirements
& $D=336$ h \citep{ArmyDirective2020_03};
$\rho^{L,N}=\rho^{PV,N}=0.15$;
$\rho^{L,A}=\rho^{PV,A}=0.16$;
$\Gamma^{L,N}=\Gamma^{PV,N}=45$; and
$\Gamma^{L,A}=\Gamma^{PV,A}=48$. \\

EDG fuel curve
& Fuel efficiency curve (load fraction, gal/h):
$(0,15.0)$, $(0.25,19.5)$, $(0.50,32.3)$, $(0.75,45.4)$, and $(1,57.3)$ \citep{CaterpillarC27_800}; convex majorant used. \\

EDG availability
& $N=13$; $q^N=0.99$; $q^A=0.999$ (study-selected entry criteria);
$p_s=0.9998$; and hourly failed-start recovery probability
$\mu=0.50$.
\\

Shifted load
& $B^{\max}=8$ MWh, power limit: 2MW;
$\eta_b=0.85$; $\beta_b=0.98$; $H^b=6$ h; and
$\beta^L=0.25$. \\

\bottomrule
\end{tabularx}
\end{table}

\subsection{Time-Varying Readiness and Dispatch Consequences}

\begin{figure}[tbp]
\centering
\includegraphics[width=\textwidth]{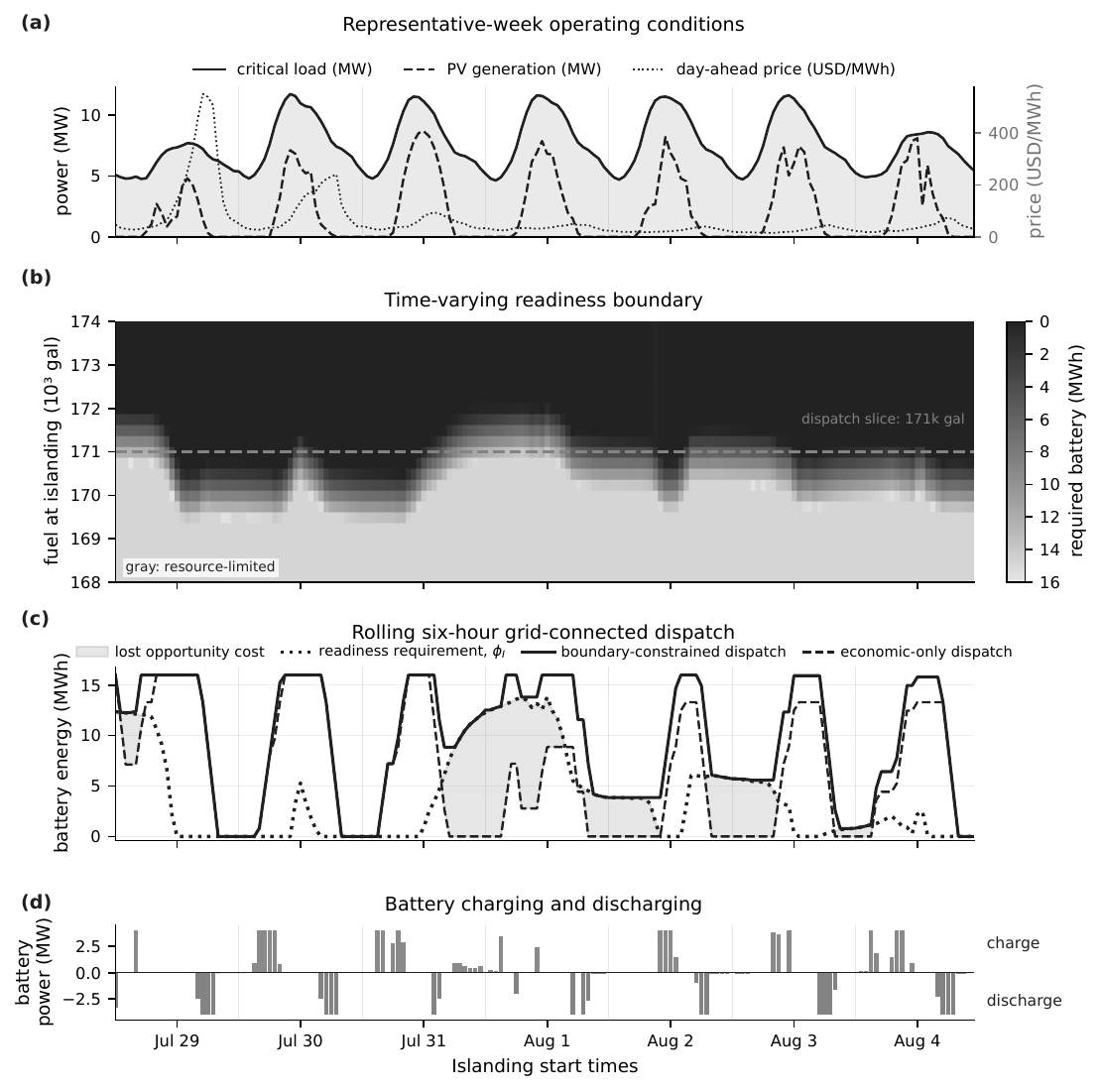}
\caption{Readiness and pre-islanding dispatch; (a) critical-load, PV, and electricity price; (b) readiness boundary over fuel inventory; (c) economic-only vs. boundary-constrained battery charge; (d) battery charge and discharge dynamics.}
\label{fig:time_varying_dispatch}
\end{figure}

Figure~\ref{fig:time_varying_dispatch} compares economic-only and readiness-constrained dispatch over the 168-hour representative week using 171,000-gal of fuel with $m=13$ and $b=0$. In the representative week, the boundary does not require any battery charge for 47 time intervals but does require battery energy for the remaining 121, with a maximum requirement of 13.79 MWh at 12:00 p.m. on August~1. Economic-only dispatch without the readiness constraint falls below the computed boundary for 78 hours, with a maximum shortfall of 12.88 MWh. Enforcing the time-varying boundary preserves the specified readiness requirement under the modeled conditions at an additional weekly cost of \$383.92, or 0.30\%, relative to economic-only operation.

\subsection{Value of Warning: Alert Reachability}

\begin{figure}[tbp]
\centering
\includegraphics[width=0.8\columnwidth]{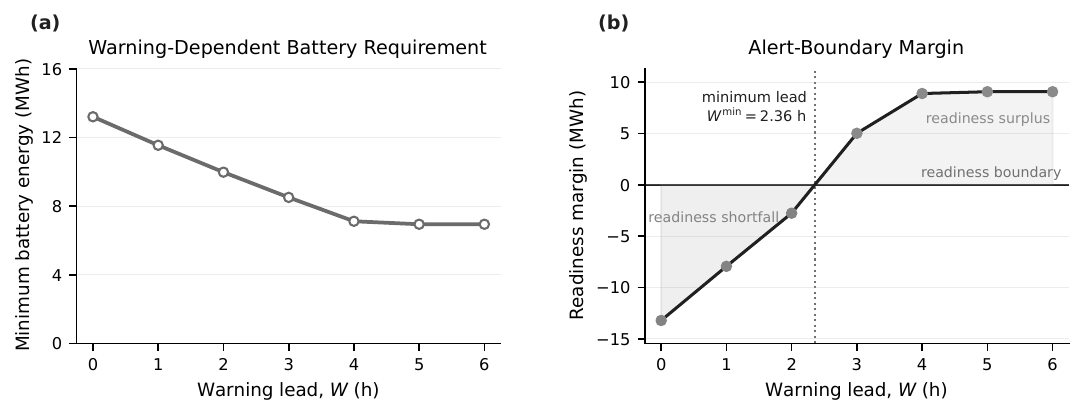}
\caption{Effect of warning on reaching alert target; (a) minimum battery energy required; (b) best attainable readiness margin}
\label{fig:warning_value}
\end{figure}

Figure~\ref{fig:warning_value} evaluates warning leads $W=0,\ldots,6$ over a common warning interval from 6:00 p.m. on July 30, 2025 to midnight at the start of July 31, 2025. Every case begins with the same readiness state: $(e,f,b,y)=(16\ \text{MWh}, 171,000\text{-gal},0,0)$. Before warning arrives, all cases follow the same least-cost normal policy. Under the alert standard $(\rho,\Gamma,q)=(0.16,48,0.999)$, the $q$-reliable EDG count is 12 across all warning leads considered. Thus, increasing $W$ provides more time for battery charging and load-shifting to reach the same alert readiness target.

Panel~(a) shows how warning changes the battery energy required at islanding. The alert boundary itself is fixed; additional warning allows more shifted load before islanding, which moves the readiness state to a point on the boundary with a lower battery-energy requirement. The requirement falls from 13.21 MWh with no warning to approximately 7 MWh after four hours of warning with strongly diminishing returns thereafter. Warning therefore reduces the amount of battery energy that must be available at islanding without relaxing the underlying readiness standard.

Panel~(b) shows the corresponding \emph{best attainable} readiness margin. With zero, one, and two hours of warning, the best attainable states remain 13.21, 7.94, and 2.76 MWh below the alert boundary, respectively. Three hours is the first evaluated lead that reaches the alert set, with a 5.01-MWh readiness surplus, but in continuous time, this boundary shift occurs at 2.36 hours. Additional warning increases the attainable surplus, but the improvement becomes small after four hours. The least-cost alert-feasible trajectory is the same for all feasible leads $W=3$--6 h and $\Delta C^A(6)=\$816.36$. Thus, the main value of warning is to make the fixed alert target reachable; once it is reachable, additional lead provides further margin in preparation of islanding.

\subsection{Comparison with Fixed Reserves and the Hussain Battery-Reserve Cut}

Table~\ref{tab:common_battery_rule_comparison} compares the proposed readiness boundary with fixed 8- and 16-MWh reserves and the $n$-interval battery cut from Eq.~(13) of Hussain et al.~\citep{hussain2018_survivability_constraint_mg_bess}, which just ensures the battery has enough charge to cover the first $n$ islanded intervals. Comparisons cover a specific hour: July~31 at 11:00 p.m., and the representative week. The single-hour cases use common forecasts and initial resources so that only the battery-energy requirement changes. The readiness premium is the operating-cost difference between readiness-constrained and economic-only operation.

At 11:00 p.m. on July~31, the 336-hour readiness boundary is 11.982 MWh. The fixed 8-MWh reserve falls below this requirement, while the 16-MWh reserve holds excess energy and incurs a larger readiness premium. This comparison illustrates the difficulty of choosing one battery reserve for all operating conditions: a smaller reserve can leave the computed readiness requirement unmet, while a larger reserve can unnecessarily restrict economic battery use. A time-varying, resource-dependent reserve addresses this tradeoff by adjusting the protected battery energy to the resources and anticipated islanding conditions.

The Hussain cut varies with forecast net load, but its battery requirement does not account for fuel, available EDGs, or load shifting. At the selected hour, the $n=1$ and $n=2$ cuts fall below the 336-hour boundary, while $n=3$ exceeds it. Across the week, both $n=2$ and $n=3$ exceed the 16-MWh battery capacity at some hours, with $n=3$ reaching 22.046 MWh. The Hussain approach easily produces infeasible battery requirements even for a reserve window of only a few hours. The proposed boundary instead accounts for the combined resources supporting the full mission when determining a sufficient battery reserve, thus allowing it to ensure readiness for multiday islanding requirements.

\begin{table}[tbp]
\caption{Fixed-Hour and Representative-Week Battery-Rule Comparison}
\label{tab:common_battery_rule_comparison}
\centering
\linespread{1}\selectfont
\small
\setlength{\tabcolsep}{2pt}
\renewcommand{\arraystretch}{1.15}

\begin{tabular*}{\textwidth}{@{\extracolsep{\fill}}lccccc@{}}
\toprule
& \multicolumn{3}{c}{July 31, 11:00 p.m.}
& \multicolumn{2}{c}{Representative Week} \\
\cmidrule(lr){2-4}\cmidrule(lr){5-6}

Readiness-Constraint Method
& \shortstack{Battery Target\\(MWh)}
& \shortstack{Readiness\\Premium (\$)}
& \shortstack{Margin\\(MWh)}
& \shortstack{Mean Readiness Premium\\(\$/case)}
& \shortstack{Shortfall Count\\(hours)} \\
\midrule

Proposed boundary
& 11.982 & 527.60 & 0.000 & 133.88 & 0 \\

Fixed 8-MWh reserve
& 8.000 & 340.45 & $-3.982$ & 409.97 & 32 \\

Fixed 16-MWh reserve
& 16.000 & 731.03 & 4.018 & 950.35 & 0 \\

Hussain, $n=1$
& 5.620 & 233.15 & $-6.361$ & 287.87 & 45 \\

Hussain, $n=2$
& 10.806 & 470.65 & $-1.176$ & 582.84$^\dagger$ & 23 \\

Hussain, $n=3$
& 15.596 & 709.29 & 3.614 & 941.63$^\dagger$ & 2 \\

\bottomrule
\end{tabular*}

\vspace{1pt}
\begin{minipage}{0.96\textwidth}
\small
$^\dagger$Certain hours require more than the physical 16-MWh battery capacity
\end{minipage}

\end{table}

\subsection{EDG Availability at Islanding and Running-Failure Stress Test}

\begin{figure}[tbp]
\centering
\includegraphics[width=\columnwidth]{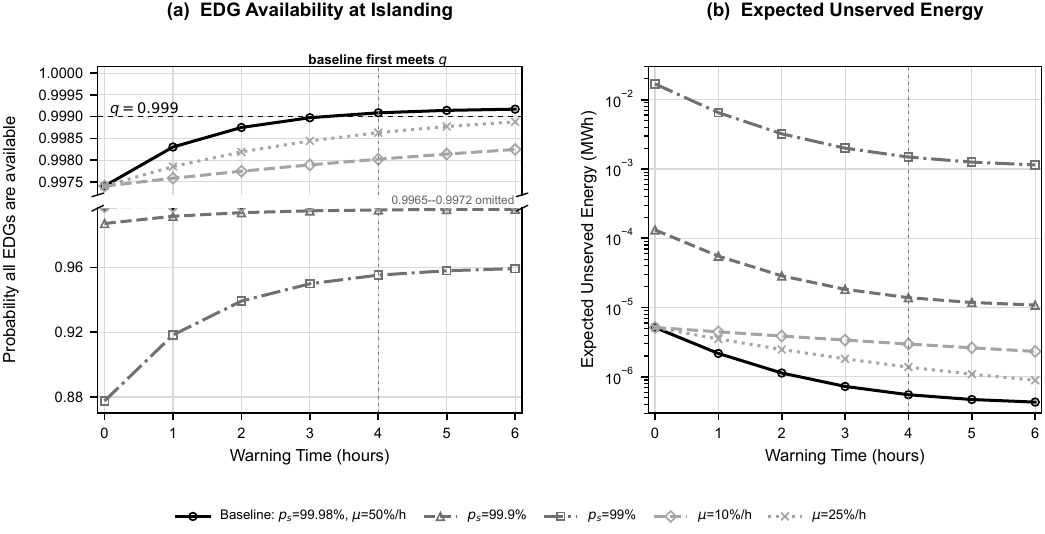}
\caption{EDG posture with warning; (a) EDG islanding-entry reliability; (b) expected unserved critical-load.}
\label{fig:edg_reliability}
\end{figure}

Figure~\ref{fig:edg_reliability} separately examines the effect of warning on EDG availability. Under the baseline probability of successful start $p_s=0.9998$ and hourly failed-start recovery probability $\mu=0.50$, the probability that all 13 EDGs are available at islanding increases from 0.997403 with no warning to 0.999088 after four hours, the first warning duration that meets the $q=0.999$ generator-entry criterion. The baseline start probability is highly optimistic and should be interpreted as an illustrative reference case rather than an estimate of typical field performance. Lower start success or slower failed-start recovery reduces the benefit of warning, and none of the four sensitivity cases reaches the all-13 requirement within the six-hour warning window. Over the parameter ranges examined, all-13 availability is more sensitive to reductions in start success than to slower failed-start recovery. These results highlight the stringent individual-generator reliability needed to achieve the all-13 availability target under the modeled preparation limits, and show that advance warning cannot necessarily compensate for lower start reliability.

The kernel excludes EDG failures after islanding begins. To examine this limitation, the case with four hours of warning is evaluated using 5,000 Monte Carlo samples along a load and PV trajectory following the uncertainty budgets. Each sample draws EDG entry outcomes and independent hourly running failures, with a simplified, deterministic six-hour recovery. Table~\ref{tab:running_reliability} reports mission survival, its 95\% Wilson interval, and expected unserved energy (EUE) for each running-failure probability. With no running failures, all samples complete the mission, but increasing failure probabilities substantially reduces survival and increases unserved energy. High EDG entry reliability alone does not ensure service throughout the mission. Commanders and installation energy personnel should therefore assess generator running reliability over the required mission duration alongside battery and fuel sufficiency, using testing and maintenance records to inform the assumed failure and recovery rates.

\begin{table}[tbp]
\caption{Ex-post running-failure stress test.}
\label{tab:running_reliability}
\centering
\linespread{1}\selectfont
\small
\begin{tabular}{@{}cccc@{}}
\toprule
Trip probability & Mission survival & 95\% interval & EUE (MWh) \\
\midrule
$0$ & 1.0000 & [0.9992, 1.0000] & 0.000 \\
$2\times10^{-4}$ & 0.9408 & [0.9339, 0.9470] & 0.358 \\
$10^{-3}$ & 0.3080 & [0.2954, 0.3209] & 36.88 \\
\bottomrule
\end{tabular}
\end{table}

\subsection{Radius and Cumulative Budget Sensitivity Analysis}

\begin{figure}[tbp]
\centering
\includegraphics[width=0.8\columnwidth]{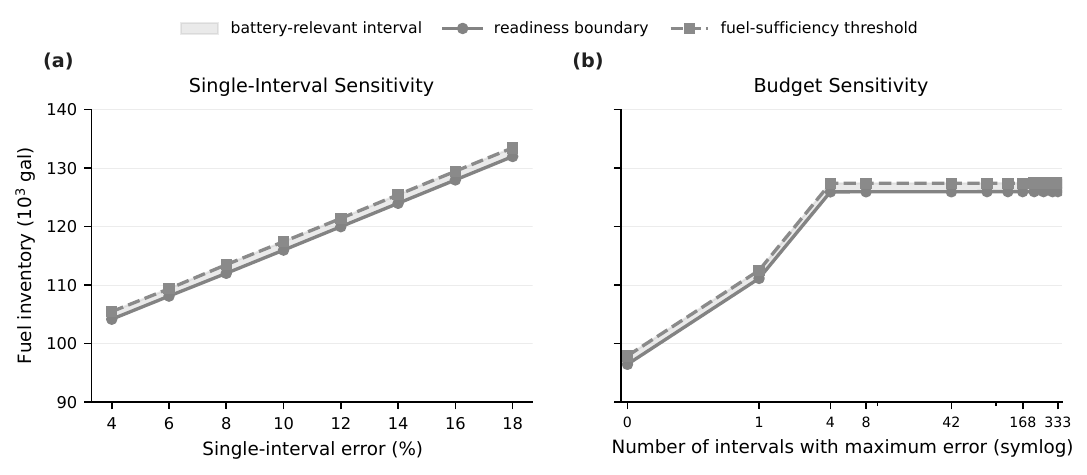}
\caption{Uncertainty severity and persistence; (a) varies error radius $\rho$ at $\Gamma/\rho=300$; (b) varies the budget in full-radius error intervals at $\rho=0.15$.}
\label{fig:uncertainty_effects}
\end{figure}

Figure~\ref{fig:uncertainty_effects} illustrates how sensitive the boundary is to different covered forecast errors and separates the roles of each constraint in the budgeted uncertainty set. Both panels report the fuel-inventory thresholds at $b=8$ MWh. At fixed $\Gamma/\rho=300$ full-radius hour-equivalents, increasing the maximum hourly forecast error from $\rho=0.04$ to 0.18 raises the readiness threshold from 104,161 to 132,010 gal and the fuel-sufficiency threshold from 105,464 to 133,464 gal. The approximately monotone 27,849-gal readiness shift shows the strong effect of single-error severity on the states that are viable.

For panel (b) at fixed $\rho=0.15$, increasing the cumulative budget from zero to one full-radius hour raises the readiness threshold from 96,455 to 111,146 gal, and allowing four full-radius hours raises it to 125,943 gal. Additional persistence has almost no effect: the readiness threshold remains within 36 gal over $\Gamma/\rho$ from 4 to 333. The same saturation appears in the fuel-sufficiency threshold. Thus the cumulative budget matters until it can cover the few most consequential intervals; beyond that point, increasing the uncertainty budget has little effect on the readiness threshold in this case.

The lower absolute fuel thresholds in this case reflect the highly varying seasonal changes in autumn load and PV conditions. Under the same configuration, the readiness threshold is 125,969 gal at midnight on October 28, compared with 169,858 gal at 7:00 p.m. on August 1. Seasonal effects are important for commanders to acknowledge because their energy readiness can vary wildly with the same amount of resources solely based on the time of year.

\subsection{Interpretation and Scope}
\label{subsec:interpretation_scope}
For installation commanders and energy managers, these results translate mission requirements into reserve and preparation decisions. Commanders and mission owners specify critical loads and required endurance, and microgrid operators compute a library of boundaries, which when queried, identify a sufficient battery reserve given available fuel, load-shifting capability, and generators. The dispatch premium quantifies the modeled cost of maintaining readiness, while the warning analysis identifies how valuable early preparation is to ensure survivability in adverse conditions.

The study also has several limitations. First, the reported readiness boundary is a conservative numerical approximation and may require more battery energy than the exact minimum. The model uses a single-bus microgrid, a fixed mission duration, known fuel inventory, and a fixed EDG count during islanding. Load and PV uncertainty parameters are operator-selected rather than statistically calibrated, so the robustness guarantee applies only to trajectories within the specified uncertainty set, but commanders and microgrid operators can collaborate to specify how to better calibrate uncertainty sets. Hourly resolution does not represent subhourly frequency, voltage, or generator-start transients. The economic analysis uses wholesale electricity prices and omits other costs like labor and maintenance. Finally, the load profile is synthetic, the PV profile uses typical-year weather, and the main operating comparison covers one week. Broader seasonal studies are needed to determine how changing load and PV conditions affect the readiness boundary.
\section{Conclusion}
\label{sec:conclusion}

This paper addresses the military operating decision of how to allocate installation energy resources during routine operations while preserving the ability to sustain mission critical loads through a prolonged loss of utility power. It develops a state-dependent alternative to fixed battery reserves for long-duration microgrid readiness. The resulting boundary is computed offline and incorporated into short-horizon grid-connected dispatch through a battery-energy readiness constraint.

Advance warning is most valuable when it closes the gap to a stronger alert target; in the simulations, 2.36 hours is the first evaluated lead that reaches the target, while additional warning primarily increases the readiness margin, with diminishing returns. The battery's C-rate limits how quickly additional energy can be stored, while its finite capacity limits further gains as the battery approaches full charge. Across the representative week, economic-only dispatch falls below the applicable readiness boundary at 78 of 168 candidate islanding times, whereas boundary-constrained dispatch has no stored-boundary violations and increases modeled weekly cost by a very modest readiness premium: \$383.92, or 0.30\%.

The readiness guarantee is conditional on the specified uncertainty set and model. Future work should extend the boundary construction to network-constrained microgrids, contested fuel logistics, and EDG failures during islanded operation. Additionally, this paper identifies how to respond to warnings of impending islanding, but more work must be done on detecting these warnings by monitoring the bulk electric grid. Broader seasonal and field studies are also needed to determine how operating conditions and uncertainty-set specification affect the readiness boundary and its economic value. The proposed framework provides operators and commanders with a method for assessing and maintaining installation energy readiness by identifying and preserving the resources needed to sustain critical missions during prolonged utility outages.

\bibliographystyle{mor}
\bibliography{references}

@ARTICLE{Broekhoven_Judson2013_military_microgrids,
  author={Van Broekhoven, Scott and Judson, Nicholas and Galvin, James and Marqusee, Jeffrey},
  journal={IEEE Power and Energy Magazine}, 
  title={Leading the Charge: Microgrids for Domestic Military Installations}, 
  year={2013},
  volume={11},
  number={4},
  pages={40-45},
  doi={10.1109/MPE.2013.2258280}}

@article{Nelson2020_microgrid_resilience,
title = {Statistical development of microgrid resilience during islanding operations},
journal = {Applied Energy},
volume = {279},
pages = {115724},
year = {2020},
issn = {0306-2619},
doi = {https://doi.org/10.1016/j.apenergy.2020.115724},
url = {https://www.sciencedirect.com/science/article/pii/S0306261920312150},
author = {James Nelson and Nathan G. Johnson and Kelsey Fahy and Timothy A. Hansen}
}

@article{marqusee2021_resilience_econ_microgrid,
title = {Resilience and economics of microgrids with PV, battery storage, and networked diesel generators},
journal = {Advances in Applied Energy},
volume = {3},
pages = {100049},
year = {2021},
issn = {2666-7924},
doi = {https://doi.org/10.1016/j.adapen.2021.100049},
url = {https://www.sciencedirect.com/science/article/pii/S266679242100041X},
author = {Jeffrey Marqusee and William Becker and Sean Ericson}
}

@Article{peterson2021_mission_impact_microgrid,
AUTHOR = {Peterson, Christopher J. and Van Bossuyt, Douglas L. and Giachetti, Ronald E. and Oriti, Giovanna},
TITLE = {Analyzing Mission Impact of Military Installations Microgrid for Resilience},
JOURNAL = {Systems},
VOLUME = {9},
YEAR = {2021},
NUMBER = {3},
eid = {69},
URL = {https://www.mdpi.com/2079-8954/9/3/69},
ISSN = {2079-8954},
DOI = {10.3390/systems9030069}
}

@techreport{judson2012_microgrid,
  author      = {Nguyen, S. V. T. and Van Broekhoven, S. B. and Judson, N. and Ross, W. D.},
  title       = {Microgrid Study: Energy Security for DoD Installations},
  institution = {MIT Lincoln Laboratory},
  type        = {Technical Report},
  number      = {1164},
  address     = {Lexington, MA, USA},
  year        = {2012},
  month       = {jun},
  url         = {https://www.ll.mit.edu/sites/default/files/publication/doc/microgrid-study-energy-security-dod-van-broekhoven-tr-1164.pdf},
}

@INPROCEEDINGS{Hutchinson2022_resilience_metrics,
  author={Hutchinson, Spencer and Bernal Heredia, Willy G. and Ghatpande, Omkar A.},
  booktitle={2022 IEEE Conference on Technologies for Sustainability (SusTech)}, 
  title={Resilience Metrics for Building-Level Electrical Distribution Systems with Energy Storage}, 
  year={2022},
  volume={},
  number={},
  pages={71-78},
  doi={10.1109/SusTech53338.2022.9794261}}

@techreport{NREL2019_microgrid_resilience_defense,
  title={Microgrids for energy resilience: A guide to conceptual design and lessons from defense projects},
  author={Booth, Samuel S and Reilly, James and Butt, Robert S and Wasco, Mick and Monohan, Randy},
  year={2019},
  institution={National Renewable Energy Laboratory, Golden, CO}
}

@article{Bertsimas2004_price_robustness,
author = {Bertsimas, Dimitris and Sim, Melvyn},
title = {The Price of Robustness},
journal = {Operations Research},
volume = {52},
number = {1},
pages = {35-53},
year = {2004},
doi = {10.1287/opre.1030.0065},
URL = { https://doi.org/10.1287/opre.1030.0065},
eprint = {https://doi.org/10.1287/opre.1030.0065}
}

@article{hussain2019_uncertainty_resilient_mg,
  title={Impact of uncertainties on resilient operation of microgrids: A data-driven approach},
  author={Hussain, Akhtar and Rousis, Anastasios Oulis and Konstantelos, Ioannis and Strbac, Goran and Jeon, Jinhong and Kim, Hak-Man},
  journal={IEEE Access},
  volume={7},
  pages={14924--14937},
  year={2019},
  publisher={IEEE}
}

@article{hussain2018_survivability_constraint_mg_bess,
  title={A proactive and survivability-constrained operation strategy for enhancing resilience of microgrids using energy storage system},
  author={Hussain, Akhtar and Bui, Van-Hai and Kim, Hak-Man},
  journal={IEEE Access},
  volume={6},
  pages={75495--75507},
  year={2018},
  doi={10.1109/ACCESS.2018.2883418},
  url={https://ieeexplore.ieee.org/document/8546726},
  publisher={IEEE}
}

@article{hussain2017_feasible_islanding_survivability,
  title={Optimal operation of hybrid microgrids for enhancing resiliency considering feasible islanding and survivability},
  author={Hussain, Akhtar and Bui, Van-Hai and Kim, Hak-Man},
  journal={IET Renewable Power Generation},
  volume={11},
  number={6},
  pages={846--857},
  year={2017},
  publisher={Wiley Online Library}
}

@book{aubin2011_viability_main_book,
  title={Viability theory: new directions},
  author={Aubin, Jean-Pierre and Bayen, Alexandre M and Saint-Pierre, Patrick},
  year={2011},
  publisher={Springer Science \& Business Media}
}

@article{saint1994_approx_viability,
  title={Approximation of the viability kernel},
  author={Saint-Pierre, Patrick},
  journal={Applied Mathematics and Optimization},
  volume={29},
  number={2},
  pages={187--209},
  year={1994},
  publisher={Springer}
}

@ARTICLE{Khodaei2014_microgrid_resilience_scheduling,
  author={Khodaei, Amin},
  journal={IEEE Transactions on Smart Grid}, 
  title={Resiliency-Oriented Microgrid Optimal Scheduling}, 
  year={2014},
  volume={5},
  number={4},
  pages={1584-1591},
  doi={10.1109/TSG.2014.2311465}}

@ARTICLE{Khodaei2014_mg_scheduling_time_islanding_constraints,
  author={Khodaei, Amin},
  journal={IEEE Transactions on Power Systems}, 
  title={Microgrid Optimal Scheduling With Multi-Period Islanding Constraints}, 
  year={2014},
  volume={29},
  number={3},
  pages={1383-1392},
  doi={10.1109/TPWRS.2013.2290006}}

@techreport{DoW2024EnergyResilience,
  author      = {{U.S. Department of War}},
  title       = {Annual Energy Performance, Resilience, and Readiness Report},
  institution = {Office of the Assistant Secretary of Defense for Energy, Installations, and Environment},
  year        = {2024},
  url         = {https://www.acq.osd.mil/eie/ero/ier/docs/aeprr/FY23-AEPRR-Report.pdf}
}

@misc{NRELOEDIEndUseProfiles,
  author       = {{National Renewable Energy Laboratory}},
  title        = {End-Use Load Profiles for the U.S. Building Stock},
  howpublished = {Open Energy Data Initiative},
  year         = {2022},
  doi          = {10.25984/1876417},
  url          = {https://data.openei.org/submissions/4520}
}

@misc{NRELPVWattsV8,
  author       = {{National Renewable Energy Laboratory}},
  title        = {{PVWatts Calculator API, Version 8}},
  year         = {n.d.},
  howpublished = {Developer Network Documentation},
  url          = {https://developer.nrel.gov/docs/solar/pvwatts/v8/},
  note         = {Accessed July 22, 2026}
}

@techreport{PJMDataMiner2,
  author      = {{PJM Interconnection}},
  title       = {Data Miner 2 API Guide},
  institution = {PJM Interconnection},
  year        = {n.d.},
  url         = {https://www.pjm.com/-/media/DotCom/etools/data-miner-2/data-miner-2-api-guide.pdf},
  note        = {Accessed July 22, 2026}
}

@manual{CaterpillarC27_800,
  author       = {{Caterpillar Inc.}},
  title        = {{C27 Diesel Generator Set: 800 ekW Standby, 60 Hz}},
  organization = {Caterpillar Inc.},
  year         = {n.d.},
  note         = {Performance No. DM7696-02; specification LEHE1213-04}
}

@techreport{ArmyDirective2020_03,
  author      = {{Secretary of the Army}},
  title       = {Army Directive 2020-03 (Installation Energy and Water Resilience Policy)},
  institution = {Department of the Army},
  address     = {Washington, D.C.},
  year        = {2020},
  month       = {March},
  url         = {https://www.asaie.army.mil/sites/default/files/PDF/ARN21689_AD2020_03_FINAL_Revised.pdf}
}

@INPROCEEDINGS{Zhang2024_proactive_weather_mg,
  author={Zhang, Yichao and Anvari-Moghaddam, Amjad and Peyghami, Saeed and Li, Yuan and Dragičević, Tomislav and Blaabjerg, Frede},
  booktitle={2024 IEEE 10th International Power Electronics and Motion Control Conference (IPEMC2024-ECCE Asia)}, 
  title={A Proactive Operating Strategy for Microgrid Resilience Enhanced for Weather-induced Outages}, 
  year={2024},
  volume={},
  number={},
  pages={4877-4882},
  doi={10.1109/IPEMC-ECCEAsia60879.2024.10567435}}
\appendix

\makeatletter\renewcommand{\@seccntformat}[1]{Appendix~\csname the#1\endcsname:\quad}
\makeatother
\section{Appendix: Numerical Boundary Construction and Algorithm}
\label{sec:offline_algorithm}

Fix $(s,D,\theta,m)$; $\widehat\phi_h(f,b,\mathbf g)$ denotes the stored battery requirement with $D-h$ intervals remaining.

\subsection{Numerical Definitions}

For the convexity argument, represent successor fuel by $f^+$ satisfying
\begin{equation}
    f^+ + \overline{r}_m\!\left(x_h^{D,L}\right)\Delta t
    \le f.
    \label{eq:convex_fuel_predecessor}
\end{equation}
With $d:=D-h$, uncertainty budget above $d\rho^j$ cannot be spent during the remaining mission. The effective budget interval for $j\in\{L,PV\}$ is
\begin{equation}
    \mathcal G_h^j
    =
    \left[
    \min\!\left\{\max\{0,\Gamma^j-h\rho^j\},d\rho^j\right\},
    \min\!\left\{\Gamma^j,d\rho^j\right\}
    \right].
    \label{eq:reachable_budget_interval}
\end{equation}

\begin{samepage}
Choose finite grids $\mathbb G_h^j\subseteq\mathcal G_h^j$, $\mathbb F_h\subseteq[0,F^{\max}]$, and $\mathbb B\subseteq[0,B^{\max}]$, including their interval endpoints, with $\mathbb G_D^j=\{0\}$. Use the same physical grid for every budget pair:
\begin{equation}
    \mathbb N_h
    :=
    \mathbb F_h\times\mathbb B
    \times\mathbb G_h^L\times\mathbb G_h^{PV}.
    \label{eq:stage_boundary_grid}
\end{equation}
\end{samepage}

For any physical resource point $(f,b)$ and gridded budget pair $\mathbf g$, set $n=(f,b,\mathbf g)$. Let $v=(\boldsymbol\zeta_v,\boldsymbol\gamma_v)\in\mathcal V_h(\mathbf g)$ denote a labeled corner constructed in Algorithm~\ref{alg:offline_boundary}. Define $\operatorname{Pred}_h(n,v;\widehat\phi_{h+1})$ as the minimum $e\in[0,E^{\max}]$ admitting an action satisfying the islanded constraints at $s+h$ under $\boldsymbol\zeta_v$ and leaving the next state at or above the stored boundary for budgets $\boldsymbol\gamma_v$. Set its value to $+\infty$ if infeasible; stop on a solver failure. The corresponding battery requirement is
\begin{equation}
    \widetilde\phi_h(n)
    =
    \max_{v\in\mathcal V_h(\mathbf g)}
    \operatorname{Pred}_h(n,v;\widehat\phi_{h+1}).
    \label{eq:nodal_predecessor_requirement}
\end{equation}

For fixed $\mathbf g$, stored requirements $u_i\ge\widetilde\phi_h(f_i,b_i,\mathbf g)$ at grid points $x_i=(f_i,b_i)$ are nonincreasing in $f,b$ and nondecreasing in either budget. The weights $\lambda_i$ in \eqref{eq:lower_convex_envelope} express $x=(f,b)$ as a convex combination of points with finite $u_i$; the minimum weighted requirement defines the interpolation.

\begin{subequations}
\begin{samepage}
\begin{align}
\mathcal I_h^{+}[u](x)
&:=
\min_{\lambda\ge0}
\left\{
    \sum_i\lambda_i u_i :
    \sum_i\lambda_i=1,\;
    \sum_i\lambda_i x_i=x
\right\},
\label{eq:lower_convex_envelope}\\
\phi^{\rm DP}_{s,h}(x,\mathbf g;m)
&\le\mathcal I_h^{+}[u](x).
\label{eq:convex_interpolation_majorant}
\end{align}
\end{samepage}
\end{subequations}
Set $\widehat\phi_h=+\infty$ when \eqref{eq:lower_convex_envelope} has no feasible weights, the physical resource bounds are violated, or the interpolated requirement exceeds $E^{\max}$. Stop if an infeasible probe has a finite stored requirement.

\subsection{Proof of Bellman Representation and Numerical Conservatism}
\label{app:bellman_proof}

\begin{proof}[Proof of Proposition~\ref{prop:bellman_certificate}]
At $h=D$, both formulations impose $\mathcal Z$. Viability at each preceding stage requires a feasible continuation after every admissible current error, exactly as in \eqref{eq:carried_budget_bellman_recursion}. Backward induction gives \eqref{eq:kernel_dp_equivalence}.

Additional fuel can remain unused, so \eqref{eq:convex_fuel_predecessor} is equivalent to the fuel-transition equality for viability. For fixed $m$, $\overline r_m$ is convex and the other constraints are affine. With a convex continuation kernel, each current error admits a convex set of feasible states. Intersecting over errors and inducting from $\mathcal Z$ proves convexity of $\mathcal K^{\rm DP}_{s,h}(\mathbf g;m)$ and its minimum-energy boundary in $(f,b)$.

For the numerical recursion, suppose $\widehat\phi_{h+1}$ is conservative; this holds at $D$. A successor state meeting this stored requirement therefore belongs to the true continuation kernel. The interpolation is convex and nonincreasing in fuel. At fixed successor budgets, \eqref{eq:convex_fuel_predecessor} and the convex continuation boundary make the predecessor feasible region jointly convex in the current errors and decision variables. Minimizing over the decision variables therefore gives a convex minimum-energy requirement, so corner values bound every error within each rectangle. Capping removes only uncertainty budget that cannot be spent during the remaining intervals. Upward rounding and componentwise maximization at shared corners can only enlarge the future uncertainty sets, so the resulting requirements remain sufficient. Consequently, the corner maximum in \eqref{eq:nodal_predecessor_requirement} covers every admissible current error and gives $u_i\ge\phi^{\rm DP}_{s,h}(x_i,\mathbf g;m)$.

For weights feasible in \eqref{eq:lower_convex_envelope}, convexity gives
\[
    \phi^{\rm DP}_{s,h}(x,\mathbf g;m)
    \le\sum_i\lambda_i\phi^{\rm DP}_{s,h}(x_i,\mathbf g;m)
    \le\sum_i\lambda_i u_i.
\]
This inequality holds for every feasible set of weights, so minimizing the right-hand side gives \eqref{eq:convex_interpolation_majorant}. Upward adjustments cannot decrease any weighted sum and therefore preserve the bound. Starting from the exact terminal boundary, backward induction establishes conservatism at every stage. At $h=0$, any state in $\mathcal Z$ meeting the stored requirement can complete the mission under every admissible uncertainty sequence, proving \eqref{eq:numerical_boundary_sufficiency}.
\end{proof}

\subsection{Offline Boundary Algorithm}

\begingroup
\singlespacing
\small
Algorithm~\ref{alg:offline_boundary} uses tolerances $\varepsilon_f>0$ for fuel spacing, $\varepsilon_e>0$ for battery energy, and $\varepsilon_{\rm num}=3\times10^{-6}$~MWh for probe checks. Interpolation uses only finite $u_i$ and returns $+\infty$ if no feasible weights exist. In refinement, $(b,\mathbf g)\in\mathbb B\times\mathbb G_h^L\times\mathbb G_h^{PV}$; equal infinite requirements have zero comparison error.

\begin{readinessalgorithm}
\readinessalgorithmcaption{Computing a conservative readiness boundary}
\label{alg:offline_boundary}
\begin{algorithmic}[1]
\Require Fixed $(s,D,\theta,m)$; $\widehat P_{s+h},\widehat L_{s+h}$; Table~\ref{tab:baseline_parameters} parameters; physical grids $\mathbb F_h,\mathbb B$ and budget grids $\mathbb G_h^L,\mathbb G_h^{PV}$
\Ensure Conservative battery-energy boundary $\widehat\phi_{s,D,\theta,m}$ at the initial uncertainty budgets
\State \textbf{Start at mission end.} $\widehat\phi_D(f,b,\mathbf 0)\gets0$, $\forall(f,b)\in[0,F^{\max}]\times[0,B^{\max}]$
\For{$h=D-1,\ldots,0$}
    \Repeat\ with $\widehat\phi_{h+1}$ held fixed
        \State \textbf{Form the grid.} $\mathbb N_h\gets\mathbb F_h\times\mathbb B\times\mathbb G_h^L\times\mathbb G_h^{PV}$
        \Statex \hspace{2em}Let $\mathbb P_h$ contain all physical grid nodes, cell-edge midpoints, and cell centers.
        \For{$\mathbf g\in\mathbb G_h^L\times\mathbb G_h^{PV}$, with each budget grid in increasing order}
            \State \textbf{Cap successor budgets.} For $j\in\{L,PV\}$ and $\zeta^j\in\Delta^j(g^j)$:
            \Statex \hspace{3em}
            $\displaystyle\gamma^j(g^j,\zeta^j)\gets\min\{g^j-\zeta^j,(D-h-1)\rho^j\}$
            \State \textbf{Round upward.}
            $\displaystyle\gamma^j(g^j,\zeta^j)\gets\min\{a\in\mathbb G_{h+1}^j:a\ge\gamma^j(g^j,\zeta^j)\}$
            \State \textbf{Partition current errors.} Divide $\boldsymbol\Delta(\mathbf g)$ into nonempty rectangles $C$ satisfying
            \Statex \hspace{3em}
            $\displaystyle\gamma^j(g^j,\zeta^j)=\gamma_C^j
            \quad\forall\boldsymbol\zeta\in C,\quad j\in\{L,PV\}$
            \State \textbf{Collect corners.} Form $\mathcal V_h(\mathbf g)$ from the corners of each $\overline C$, paired with its budgets:
            \Statex \hspace{3em}
            $\displaystyle v=(\boldsymbol\zeta_v,\boldsymbol\gamma_v),\qquad
            \gamma_v^j\gets\gamma_C^j,\quad j\in\{L,PV\}$
            \State \textbf{Merge shared corners.} At each shared $\boldsymbol\zeta_v$, retain the larger budget in each component:
            \Statex \hspace{3em}
            $\displaystyle\gamma_v^j\gets
            \max\{\gamma_C^j:\boldsymbol\zeta_v\in\overline C\},
            \quad j\in\{L,PV\}$
            \For{$(f,b)\in\mathbb F_h\times\mathbb B$}\label{alg:boundary_nodes}
                \State \textbf{Compute battery requirement.}
                \Statex \hspace{4em}
                $\displaystyle\widetilde\phi_h(f,b,\mathbf g)\gets
                \max_{v\in\mathcal V_h(\mathbf g)}
                \operatorname{Pred}_h((f,b,\mathbf g),v;\widehat\phi_{h+1})$
            \EndFor
            \State \textbf{Initialize stored values and constraints.} $u_i\gets\widetilde\phi_h(f_i,b_i,\mathbf g)$; $\mathcal C\gets\varnothing$
            \begin{samepage}
            \State \textbf{Evaluate off-grid probes.} For $x=(f,b)\in\mathbb P_h\setminus(\mathbb F_h\times\mathbb B)$:
            \Statex \hspace{3em}
            $\displaystyle\widetilde\phi_h(x,\mathbf g)\gets
            \max_{v\in\mathcal V_h(\mathbf g)}\operatorname{Pred}_h((x,\mathbf g),v;\widehat\phi_{h+1})$
            \Repeat\label{alg:boundary_lift}
                \State \textbf{Adjust stored values.} Minimize $\sum_i u_i$ over finite $u_i$, subject to
                \Statex \hspace{4em}$u_i\ge\widetilde\phi_h(x_i,\mathbf g)$, nonincrease in $f,b$, nondecrease in budgets, and all constraints in $\mathcal C$.
                \State \textbf{Interpolate probes.} For $x\in\mathbb P_h$, compute $q(x)$ and optimal weights $\lambda^*(x)$, if any:
                \Statex \hspace{4em}
                $\displaystyle q(x)\gets\min_{\lambda\ge0}
                \left\{\sum_i\lambda_i u_i:\sum_i\lambda_i=1,\;\sum_i\lambda_i x_i=x\right\}$
                \State \textbf{Identify violations.}
                \Statex \hspace{4em}
                $\displaystyle\mathcal A\gets\left\{x\in\mathbb P_h:
                \widetilde\phi_h(x,\mathbf g)<+\infty,\quad
                q(x)+\varepsilon_{\rm num}<\widetilde\phi_h(x,\mathbf g)\right\}$
                \State \textbf{Retain constraints.} Add $\sum_i\lambda_i^*(x)u_i\ge\widetilde\phi_h(x,\mathbf g)$ to $\mathcal C$, $\forall x\in\mathcal A$.
            \Until{$\mathcal A=\varnothing$}
            \end{samepage}
            \State \textbf{Interpolate the boundary.}\label{alg:boundary_fit}
            \Statex \hspace{3em}
            $\displaystyle\widehat\phi_h(f,b,\mathbf g)\gets
            \min_{\lambda\ge0}\left\{\sum_i\lambda_i u_i:
            \sum_i\lambda_i=1,\;\sum_i\lambda_i x_i=(f,b)\right\}$
            \Statex \hspace{3em}Set $+\infty$ outside resource bounds or above $E^{\max}$; stop if an infeasible probe has finite $\widehat\phi_h$.
        \EndFor
        \State \textbf{Initialize refinement.} $\mathbb M_h\gets\varnothing$
        \For{consecutive $f<f'$ in $\mathbb F_h$ with $f'-f>\varepsilon_f$}
            \State \textbf{Form midpoint.} $\bar f\gets\tfrac12(f+f')$
            \State \textbf{Evaluate midpoint.} For every $(b,\mathbf g)$:
            \Statex \hspace{3em}
            $\displaystyle\widetilde\phi_h(\bar f,b,\mathbf g)\gets
            \max_{v\in\mathcal V_h(\mathbf g)}
            \operatorname{Pred}_h((\bar f,b,\mathbf g),v;\widehat\phi_{h+1})$
            \State \textbf{Compute error.} $\delta\gets\max_{b,\mathbf g}\bigl|\widehat\phi_h(\bar f,b,\mathbf g)-\widetilde\phi_h(\bar f,b,\mathbf g)\bigr|$
            \If{$\delta>\varepsilon_e$ \textbf{or} $\widehat\phi_h$ has different endpoint regimes for some $(b,\mathbf g)$}
                \State $\mathbb M_h\gets\mathbb M_h\cup\{\bar f\}$
            \EndIf
        \EndFor
        \State \textbf{Refine the grid.} $\mathbb F_h\gets\mathbb F_h\cup\mathbb M_h$
    \Until{$\mathbb M_h=\varnothing$}
\EndFor
\State \Return $\widehat\phi_{s,D,\theta,m}(f,b)
    :=\widehat\phi_0(f,b,(\Gamma^L,\Gamma^{PV}))$
\end{algorithmic}
\readinessalgorithmend
\end{readinessalgorithm}
\par
\endgroup

\end{document}